\documentclass[10pt,conference]{IEEEtran}

\def\BibTeX{{\rm B\kern-.05em{\sc i\kern-.025em b}\kern-.08em
    T\kern-.1667em\lower.7ex\hbox{E}\kern-.125emX}}

\usepackage[numbers]{natbib}

\usepackage{float}

\usepackage[utf8]{inputenc}

\usepackage{csquotes}

\usepackage{graphicx}
\graphicspath{{images/}}

\PassOptionsToPackage{hyphens}{url}\usepackage{hyperref}

\usepackage{array}

\usepackage{tabularx}

\usepackage{multirow}

\usepackage{xcolor}
\definecolor{deepblue}{rgb}{0,0,0.5}
\definecolor{deepred}{rgb}{0.6,0,0}
\definecolor{deepgreen}{rgb}{0,0.5,0}
\definecolor{commentgreen}{RGB}{2,112,10}
\definecolor{eminence}{RGB}{108,48,130}
\definecolor{weborange}{RGB}{255,165,0}
\definecolor{frenchplum}{RGB}{129,20,83}

\usepackage{listings}
\usepackage{pythonhighlight}
\begin{document}

\title{
    A Case Study on AI Engineering Practices: Developing an Autonomous Stock Trading System
}

\author{
    \IEEEauthorblockN{Marcel Grote}
    \IEEEauthorblockA{
        University of Stuttgart, Institute of Software Engineering\\
        Stuttgart, Germany\\
        st161027@stud.uni-stuttgart.de
    }
    \and
    \IEEEauthorblockN{Justus Bogner}
    \IEEEauthorblockA{
        University of Stuttgart, Institute of Software Engineering\\
        Stuttgart, Germany\\
        justus.bogner@iste.uni-stuttgart.de
    }
}

\maketitle


\begin{abstract}
Today, many systems use artificial intelligence (AI) to solve complex problems.
While this often increases system effectiveness, developing a production-ready AI-based system is a difficult task.
Thus, solid AI engineering practices are required to ensure the quality of the resulting system and to improve the development process.
While several practices have already been proposed for the development of AI-based systems, detailed practical experiences of applying these practices are rare.

In this paper, we aim to address this gap by collecting such experiences during a case study, namely the development of an autonomous stock trading system that uses machine learning functionality to invest in stocks.
We selected 10 AI engineering practices from the literature and systematically applied them during development, with the goal to collect evidence about their applicability and effectiveness.
Using structured field notes, we documented our experiences.
Furthermore, we also used field notes to document challenges that occurred during the development, and the solutions we applied to overcome them.
Afterwards, we analyzed the collected field notes, and evaluated how each practice improved the development.
Lastly, we compared our evidence with existing literature.

Most applied practices improved our system, albeit to varying extent, and we were able to overcome all major challenges.
The qualitative results provide detailed accounts about 10 AI engineering practices, as well as challenges and solutions associated with such a project.
Our experiences therefore enrich the emerging body of evidence in this field, which may be especially helpful for practitioner teams new to AI engineering.
\end{abstract}


\section{Introduction}
Today, more and more critical software systems are based on artificial intelligence (AI) and machine learning (ML), such as autonomous cars, power grid management software, or autonomous stock trading systems~\cite{dietterich2017steps}.
While this enables functionality that was previously impossible, AI-based systems also come with additional complexity and new engineering challenges~\cite{Martinez-Fernandez2022}.
Examples are efficiently managing large amounts of data~\cite{lwakatare2019taxonomy}, ensuring system safety and reliability~\cite{varshney2016engineering,gula2020software}, managing new types of technical debt~\cite{Bogner2021}, or choosing the right architecture~\cite{Serban2022}. 
Because the failure of a critical system can have serious negative consequences, it is important to ensure the quality of the developed system~\cite{EvolvingCriticalSystems}.
To achieve this, software engineering research and practice have proposed numerous guidelines and best practices for the development of conventional software systems, and their usage has increased substantially over the last 40 years~\cite{zelkowitz1984software,garousi2015survey}.
However, comparatively fewer works have tried to do the same regarding AI engineering practices.
Examples are quality assurance guidelines by \citet{hamada2020guidelines}, best practices distilled from industry experience at Microsoft~\cite{amershi2019software}, and best practices synthesized from white and gray literature by \citet{serban2020adoption}.
With the formation of AI engineering practices, questions about their effectiveness and concrete usage are starting to arise, e.g., when and how should which practices be applied?
So far, very few publications report concrete experiences of applying these practices, as well as their effects or potential challenges for using them.

In this qualitative study, we therefore aim to provide additional evidence for the applicability and effectiveness of AI engineering practices.
We accomplish this by applying 10 proposed practices during a case study, namely the development of a concrete AI-based system that predicts stock prices using ML models and autonomously performs trades based on the results.
We show how applying each of the 10 practices can improve the development and where it has its place in the ML development process.
Furthermore, we synthesize generalizable insights on experienced challenges during the development and solutions to overcome them.
We hope that our results can support AI engineering teams during the adoption of these practices, and ultimately enable them to apply the practices more effectively.
Similarly, the experienced and solved challenges can be especially helpful for newcomers to the field of AI engineering.

\section{Background and Related Work}
In this section, we explain fundamental concepts necessary to understand this study and highlight related work in the area.

\subsection{AI-based Systems and AI Engineering}
To minimize risk and to ensure that a system fulfills all requirements, it is common to follow proven software engineering practices.
In 1968, Fritz Bauer described the term software engineering as \enquote{the establishment and use of sound engineering principles in order to obtain economically software that is reliable and works efficiently on real machines}~\cite{martin1988structured}.
Since then, the field of software engineering has evolved massively~\cite{Boehm2006}.
The increasing usage of AI techniques such as machine learning introduced substantial changes, as the development of AI-based systems differs from the development of traditional software~\cite{Ozkaya2020}.
\citet{amershi2019software} identified three fundamental differences: 1) data discovery, management, and versioning are more complex, 2) development requires a wider set of skills, and 3) achieving a modular design is more difficult, since AI components can be entangled in complex ways.
These differences create the need for new practices for the development and evolution of AI-based systems~\cite{Martinez-Fernandez2022}.

To update and adapt software engineering practices for this new context, the field of \textit{AI Engineering} is starting to form~\cite{Bosch2021}.
For example, the US Office of the Director of National Intelligence (ODNI) funded an initiative to advance the discipline of AI engineering for defense and national security at the Carnegie Mellon Software Engineering Institute (SEI)~\cite{seisponsoring}.
The SEI has defined AI engineering as \enquote{a field of research and practice that combines the principles of systems engineering, software engineering, computer science, and human-centered design to create AI systems in accordance with human needs for mission outcomes}~\cite{ozkaya2021ai}.

\subsection{Autonomous Stock Trading Systems}
Autonomous systems are capable to perform unsupervised operations and to make decisions without human intervention~\cite{Sifakis2019}.
However, this lack of human oversight creates additional challenges for the quality assurance of such systems, e.g., regarding functional correctness, safety, and fairness.
The integration of ML components into an autonomous system for prediction functionality is a particularly powerful combination~\cite{zhang2017current}.
One popular example we selected for this case study is to use ML to predict stock price movements to support stock trading.
There are numerous papers on feasible machine learning models for this use case~\cite{ml4trading3,ml4trading2,ml4trading1}, which makes it especially interesting to take a more holistic, system-centric perspective on this topic. 
Furthermore, developing such an autonomous stock trading system is complex enough to not be discounted as a \enquote{toy example} and challenging enough to require sound engineering practices.
There are many different investment strategies for such a system.
They can be broken down according to how long the investor intends to hold the stock before selling it.
In this study, we mainly consider holding the investment for less than a day.
This is described as \textit{intraday trading}.
In some cases, we also consider holding the investment for up to one week, which is referred to as a \textit{short-term investment}.

\subsection{Related Work}
Several existing publications discuss AI engineering practices.
\citet{serban2020adoption} identified 29 engineering practices to develop machine learning systems by conducting a literature review.
Additionally, they conducted a survey with 313 practitioners to identify the degree of adoption of these practices.
They concluded that some practices should receive more attention, while some other practices should receive less.

\citet{akkiraju2020characterizing} present 33 best practices from the authors' personal experiences for different stages of the life cycle of an ML model.
Using the capability maturity model (CMM) as inspiration, they constructed a maturity framework in which they organize these practices to help organizations achieve a higher level of maturity for AI application development.

\citet{Nguyen-Duc2020} conducted a multiple case study in seven companies to find out how AI-based systems are developed and what software engineering processes and practices can be applied.
They analyzed the respective context for AI system development, and paid special attention to how the companies related business opportunities to AI systems.
They concluded that AI engineering practices and especially their adoption in companies would still be poorly understood.

Similarly, Ashiku et al.~\cite{ashiku2021machine} conducted a case study in which they used tools to develop ML models with the aim of evaluating the acceptance of transplanted kidneys.
They analyzed these tools mainly based on how much they simplified the development of the corresponding model.
Although we used certain ML-related tools in our case study, our focus was more on more general practices and a system perspective.

Additionally, several studies have examined challenges during the development of AI-based systems.
\citet{lwakatare2019taxonomy} conducted a multiple case study to identify the challenges companies face during the development of machine learning systems.
The challenges were mapped onto a taxonomy based on the different evolution stages of using ML components.
They concluded that a lot of effort is needed to manage those challenges that they identified as the most important ones.

In a similar study, \citet{de2019understanding} conducted interviews with developers from three companies developing ML systems.
From a process perspective, they identified the main challenges and proposed checklists for mitigation.
Afterwards, they evaluated the checklists with a focus group.

Lastly, \citet{l2017machine} summarized big data challenges for machine learning model development.
They also presented approaches to overcome these challenges and state that, with ever-growing data sets, addressing these challenges becomes increasingly important.

In summary, several articles on AI engineering practices and challenges exist, but only very few are concerned with their concrete applicability and detailed, practical experiences.
Existing work mostly focuses exclusively on either practices or challenges, and often takes a fairly model-centric perspective.
Additionally, most publications abstractly mention how much these practices or solutions simplify the development of the corresponding machine learning model, but usually ignore the concrete context of applying the practice and the resulting effects on software quality.

\begin{figure*}
    \centering
    \includegraphics[width=\textwidth]{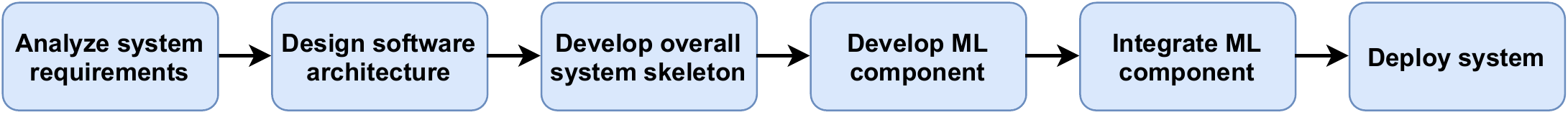}
    \caption{Development process during the case study}
    \label{fig:developmentProcess}
\end{figure*}

\section{Study Design}
We therefore want to provide rich, qualitative evidence about the applicability and effect of AI engineering practices in the context of developing a concrete AI-based system.
This research was performed as a case study~\cite{runeson2009guidelines} and guided by the following three research questions.

\textbf{RQ1:} How effective are the proposed practices regarding the development of an AI-based system?

Many AI engineering practices have been proposed.
Since it is difficult to select which ones to focus on first, it is necessary to collect experiences on how effective these practices are, and when and how they are best applied.
During this study, we applied a selection of 10 proposed practices and analyzed the context and effectiveness of each practice.

\textbf{RQ2:} What challenges can occur during the development of an AI-based system?

Software development contains many potential challenges in general, but developing software with machine learning components comes with a variety of new ones.
We identified and analyzed the challenges that occurred during the case study and compared them to existing literature.

\textbf{RQ3:} How can the challenges be addressed?

Additionally, we attempted to overcome the experienced challenges in the best possible way.
We described our chosen solutions and compared them to possible alternatives, with the goal of identifying the most beneficial ones regarding the required time and effects on software quality.

\subsection{Case Description}
We conducted a \textit{holistic} case study~\cite{runeson2009guidelines} with a single unit of analysis, namely the development of an autonomous stock trading system.
The system uses machine learning functionality to predict upward movements in stock prices.
This information is then used to autonomously trade stocks.
The general objective of the system is to make a profit through its investments.
Before beginning the development, we surveyed the literature for suitable AI engineering practices to apply.
Any encountered challenges and the usage and effect of each practice were documented and analyzed afterwards.
Note that the goal was \textit{not} to develop the best, most profitable system possible, but to use this realistic case as a vessel to apply and evaluate AI engineering practices.

The development took three months and followed a fairly standard software engineering process (see Fig.~\ref{fig:developmentProcess}).
We started with an analysis of the system requirements, followed by a design stage and then the implementation.
Most implementation tasks were carried out by one developer, the first author, with the remaining authors engaging in discussion and providing feedback.
The implementation began with the development of the overall system skeleton and ended with the development and fine-tuning of the machine learning component.
After ML component development was finished, the result was integrated into the overall system during the integration stage.
Finally, the system was deployed.
While this process appears fairly linear, several small iterations occurred, especially after meetings with external stakeholders.

\subsection{Selection of AI Engineering Practices}
Based on our survey of related work, we selected two publications that offered the most holistic set of practices when combined.
These practices affect the entire development process of an AI-based system with ML components, ranging from requirements analysis up to deployment and maintenance.
\citet{serban2020adoption} accumulated 29 best practices and grouped them into 6 categories, namely \textit{data}, \textit{training}, \textit{deployment}, \textit{coding}, \textit{team}, and \textit{governance}.
\citet{akkiraju2020characterizing} presented 33 best practices, which we grouped into the same categories as \citeauthor{serban2020adoption}.
In total, we had a set of 62 practices to choose from, even though several practices were very similar or had considerable overlap.
To keep the number of evaluated practices manageable, we selected 10 practices for our case study.
We aimed for practices to improve quality in several categories, e.g., \textit{data}, \textit{training}, \textit{deployment}, etc., but also picked some from which we expected an effect on the majority of the development process.
The following practices were selected:

\begin{enumerate}
    \item \textbf{Standardize and automate data quality check procedures} to ensure that only valid data is used for training or testing~\cite{akkiraju2020characterizing}.
    \item \textbf{Use error validation and categorization} to provide insights into when and why the ML model fails so that its reliability can be improved~\cite{akkiraju2020characterizing}.
    \item \textbf{Capture the training objective in a metric that is easy to measure and understand} to increase interpretability and to avoid entangled measurements~\cite{serban2020adoption}.
    \item \textbf{Use cross-validation} to avoid testing an ML component on data that it has already seen~\cite{akkiraju2020characterizing}.
    \item \textbf{Continuously measure model quality, performance, and drift} to detect and fix errors early~\cite{serban2020adoption,akkiraju2020characterizing}.
    \item \textbf{Review model training scripts} to ensure their quality~\cite{serban2020adoption}.
    The original practice proposes peer review.
    Since most of the development in our case study was carried out by one developer, we generalized this practice to allow a later review by the same developer.
    \item \textbf{Test all feature extraction code} to ensure that the transformed data is consistent and accurate~\cite{serban2020adoption}.
    \item \textbf{Automate hyperparameter optimization and model selection} to save exploration time and increase model quality~\cite{serban2020adoption}.
    \item \textbf{Log prediction results together with model version and input data} to provide insights into how the model can be improved~\cite{serban2020adoption}.
    The original practice proposed to log predictions after deployment.
    Because we expected this practice to be useful elsewhere, we generalized it to be also used during development and testing.
    \item \textbf{Collaborate with multidisciplinary stakeholders} to simplify the development and improve the resulting software via domain-specific knowledge~\cite{serban2020adoption}.
    The original practice only considers collaborating with team members, but we generalize this practice to additionally include relevant external stakeholders.
\end{enumerate}

\subsection{Data Collection}
The primary means for data collection during our case study were structured \textit{field notes}~\cite{Seaman1999}.
Before the development, we considered at what moments during the development each practice might be most effective.
When we reached one of these points, we then applied the corresponding practice and subsequently documented the experience using a field note template.
Each note contained the corresponding date, the title of the applied practice, a description of how it was applied, a textual description of its perceived effects, and a subjective effectiveness score on a 5-point ordinal scale ($--$, $-$, $0$, $+$, $++$).
This scale represented a rating of how much a practice (in our opinion) had improved the system or development process in this instance.
It contained the following labels:
\begin{itemize}
    \item very ineffective: $--$
    \item ineffective: $-$
    \item neutral (neither ineffective nor effective): $0$
    \item effective: $+$
    \item very effective: $++$
\end{itemize}
In cases where we could easily measure the impact of a practice, e.g., as an improvement of the model precision, we additionally documented these measurements.
For any performance measurements, we used a laptop with an Intel Core i7-8550U with four physical cores with a clock rate of 1.8 GHz, 16 GB RAM, Windows 10 21H1 64-bit as operating system, and Python 3.10.
In total, 28 field notes on the application of practices were accumulated.

Similarly, we used another field note template to document challenges that occurred during the development.
These notes contained a description of the challenge, the date of occurrence, the source that made us aware of it, e.g., personal experience or a certain stakeholder, and a description of the solution including its perceived effectiveness.
A challenge was documented as soon as it was consciously experienced.
After overcoming a challenge, we completed the corresponding field note by adding the applied solution.
For challenges that occurred more than once, we also aggregated the frequency.
In total, we collected 34 field notes on challenges during the case study, which we synthesized into 9 AI-specific challenges.

\subsection{Data Analysis}
To answer RQ1, we aggregated and interpreted the corresponding field notes on practices.
Based on the descriptions and effectiveness ratings, we synthesized how much applying each practice or tool had simplified the development process and how much it had affected software quality attributes such as maintainability, model accuracy, or performance efficiency.
Less clear results were discussed within the research team to reach a conclusion.
In addition, we compared our results with existing literature on these practices.

For RQ2, we aggregated and analyzed the field notes on  challenges.
As with the practices, we also compared the results with existing literature on AI engineering challenges.

To answer RQ3, we analyzed the documented applied solutions and synthesized how easy it was to apply them, how effective they solved the corresponding challenge, and the effects on software quality.
In some cases, we compared the applied solution with alternatives that were also considered.
Lastly, we contrasted our solutions with existing literature.
For transparency and replicability, we share our artifacts online.\footnote{\url{https://doi.org/10.5281/zenodo.7566146}}

\begin{figure*}
    \centering
    \includegraphics[width=\textwidth]{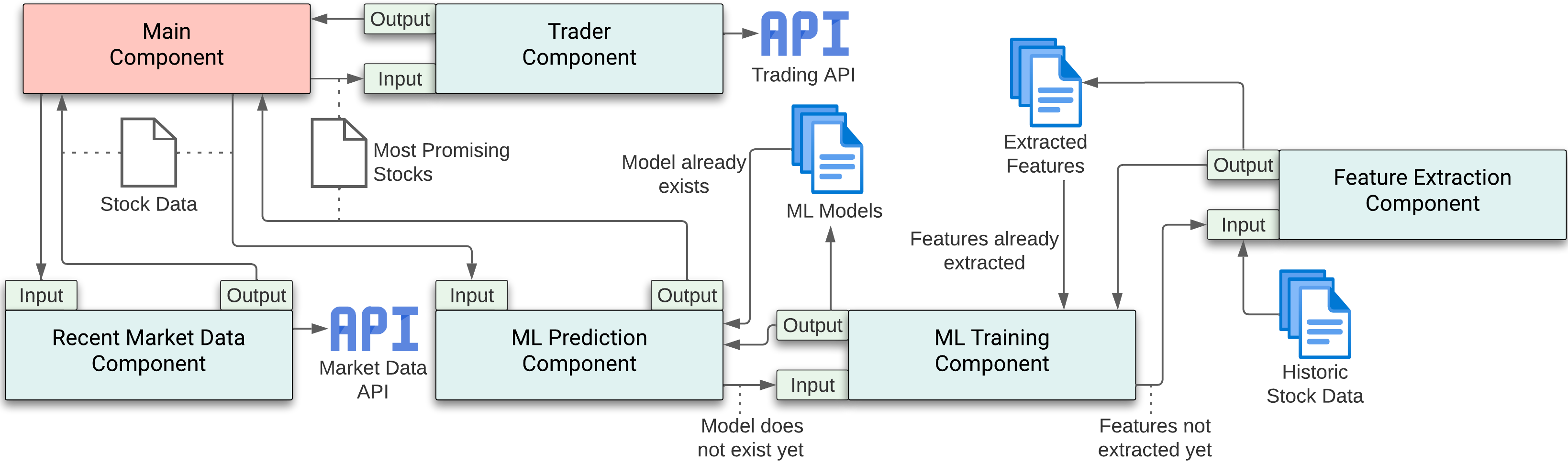}
    \caption{Diagram of the system architecture}
    \label{fig:arch}
\end{figure*}

\section{Results}
In this section, we first give a brief overview of the developed system and its architecture.
We then present how effective the applied practices were during development (RQ1).
For RQ2 and RQ3, we then describe encountered challenges and explain how we addressed them.

\subsection{Autonomous Stock Trading System}
The trading system\footnote{\url{https://github.com/Marcel0503/Autonomous-Stock-Trading-System}} was programmed exclusively in Python. 
It consists of several modularized Python files that are called from a main file (see Fig.~\ref{fig:arch}).
For intelligence creation, the \texttt{ML Training} component trains an ML model using a large data set of historic stock data.
During this process, the \texttt{Feature Extraction} component is used to provide the input for training.
The final model is embedded into the \texttt{ML Prediction} component, which corresponds to the intelligence implementation of the system.
This component is called from the \texttt{Main} component with data received via the \texttt{Recent Market Data} component, which in turn got the data from an external API.
The results of this inference, namely the most promising stocks, are then handed over to the \texttt{Trader} component, which again uses the external API to invest in these stocks.

Pandas\footnote{\url{https://pandas.pydata.org}} is used to store extracted features, while historical stock data for ML training is stored in CSV files.
For each stock, each historic day represents one data point consisting of seven features: relative difference between highest and lowest price on the previous day, relative difference between closing and opening price, moving averages of the price over the last 7, 14, and 21 days, standard deviation divided by moving average over the last 7 days, and lastly, stock volume on the previous day.
Additionally, each data point is labeled according to the highest stock price improvement on that day.
Before the stock market opens, the running system uses the Alpaca API\footnote{\url{https://alpaca.markets}} to receive stock data for the last 21 days to create data points in the training data format.

A set of k-nearest-neighbor models predict the ranking of most profitable stocks for the current day.
As soon as the stock market opens, the system buys shares of the top five stocks and sets a take-profit price based on the predictions plus a predefined stop-loss price.
Monitoring the investments is not necessary, as the trading platform automatically sells the bought shares once the take-profit or stop-loss price is reached.
Additionally, the system sells all remaining shares three minutes before the stock market closes.
This avoided a large number of API calls over the day.
We decided that the system should invest at the Nasdaq stock market\footnote{\url{https://www.nasdaq.com}}, since this is the largest global electronic market for trading stocks~\cite{nasdaqglobal}.

\subsection{Effectiveness of AI Engineering Practices (RQ1)}
For each practice, we first present how it was applied and then discuss its perceived effectiveness (\textit{analysis}).
A summary is presented in Table~\ref{table:practice-summary}.

\textbf{Standardize and automate data quality check procedures:}
We applied this practice twice.
First, we used it to validate retrieved intraday stock data.
The retrieval script waited 12 seconds between requests to avoid API rate limiting.
Data for each stock was split into 24 files, which at first were checked manually.
When this became infeasible, we extended the retrieval script with automatic quality checks.
The script checked for missing files and retried the retrieval in this case.
Additionally, it checked for files with less than 100 lines, as this indicated erroneous data.
Such files were deleted to avoid training with corrupted data.
Both types of events were also logged with an error message.

Later in the development, we also retrieved aggregated daily stock data.
To avoid missing days, we implemented another quality check.
Closed days like weekends and public holidays needed to be respected, so we compared gaps between the different stocks.
This procedure revealed that 17 of the 716 stocks contained real gaps.
In total, 137 days were missing.
Since for many stocks only one or two days were missing, we decided to still use them as training data and only ignored stocks missing more than two consecutive days.
The biggest gap we discovered was over a period of 28 days.

\textit{Analysis:}
Since we received more than 9,000 files, automated procedures were critical to identify data sets that were likely to have a negative effect on ML training.
The practice was fairly effective, as it helped us to detect many gaps in the intraday and daily trading data.
However, it would have been very time-consuming to ensure near perfect data quality.
This would have required at least a second data source for comparison.
Nonetheless, without this practice, we would have only been able to detect a fraction of the existing data quality issues, and even the most sophisticated ML algorithms perform poorly when trained with bad data~\cite{whang2020data}.
Applying this practice definitely improved the development process.

\textbf{Use error validation and categorization:}
As proposed by Akkiraju et al.~\cite{akkiraju2020characterizing}, we used a confusion matrix to identify training errors.
Such a visualization of correct and incorrect predictions easily shows in which areas a model is not effective.
For example, we identified that a decent number of our incorrect predictions were false negatives.
Although this was not ideal, it was also less severe than false positives: false negative predictions would make the system only miss out on good investments, but it would not lose any money.
As a result, we categorized these errors as less important than false positives, since these would make the system take unprofitable investments.
Therefore, we focused on reducing the number of false positives and optimized for precision.

Another training error that occurred many times was underfitting, mostly due to using too many features.
Some initial models were trained with more than 60 features, which led to a poor confusion matrix.
Additionally, we detected some minor flaws, e.g., that the model did not consider any market correlation.
If the stock price of most car companies is massively decreasing, the system should probably not invest in any car companies during this week.
Since this was complex to implement, we categorized this issue as too costly to fix.

\textit{Analysis:}
Akkiraju et al.~\cite{akkiraju2020characterizing} proposed that by prioritizing errors based on the severity and business value of fixing them, it is possible to improve the model more efficiently.
This was exactly the case in our study.
By prioritizing truly costly errors, we avoided wasting time fixing errors that barely impacted system effectiveness.
Therefore, we were able to develop a more effective system in a more efficient way.

\textbf{Capture the training objective in a metric that is easy to measure and understand:}
We first defined the training objective for the ML model to predict if a stock increases by at least 1.6\% during the next day with at least 90\% accuracy.
However, these initial models were ineffective, and we had to redefine our objective.
This time, we decided to use the overall system-level objective to invest profitably in stocks.
We experimented with multiple corresponding low-level training objectives, e.g., predicting if stock prices increase in the next five days by a certain threshold.
Furthermore, the model precision required to make the system profitable also varied depending on the approach.
We generally defined that the precision had to be high enough for the system to be profitable, even though this is difficult to measure during training and is somewhat in conflict with the proposed practice.

The deployed system uses 19 different models.
For each model, we defined the training objective as predicting whether a stock price will increase above a specific threshold.
Additionally, we simulated how much profit the entire system would make using the predictions of all 19 models and used the results of this simulation to identify whether the general objective of making a profit is achieved.

\textit{Analysis:}
Using the practice resulted in helpful reflection about our system objectives and improved communication.
During a meeting with the ML engineer, we could easily explain the system objective and the individual model objectives.
Simplifying communication during the development would be even more helpful when working in a team with many developers.
However, the training objective may also change during development, and ensuring that all team members are aware of this is important.
Additionally, there are cases where it is very difficult to capture the system objective in a metric that is easy to measure.
We had to develop a complex simulation to estimate if the objective would be achieved.

\textbf{Use cross-validation:}
We tested models using \textit{holdout cross-validation}, which splits the entire data randomly into two mutually exclusive data sets~\cite{holdoutexplain}, one for training the model and one for testing it.
This ensured more reliable results because the model was tested only on data outside the training set.
For illustration, we trained and validated a random forest model for our data once with cross-validation and once without it.
The cross-validation variant achieved a precision of 65\%, whereas the variant without it achieved over 99\%.
Testing the model on data that it has already seen during training clearly produces results that cannot be trusted.
Furthermore, we used an efficient implementation of the \textit{leave-one-out cross-validation} to test a k-nearest-neighbor model.
This procedure uses all data points for training except one, which is used for validation~\cite{leaveoneoutexplain}.
This is then repeated with each data point, which is time-intensive, but allows validating the model on more data.
We additionally aimed to identify how much cross-validation affects the test results.
We tested three different k-nearest-neighbor models with $k = 5, k = 50$, and $k = 100$.
Each model was validated once without cross-validation and once using leave-one-out cross-validation.

\textit{Analysis:}
In our study, the effect of using cross-validation varied strongly depending on the model.
For the model with $k = 5$, not using cross-validation falsely indicated a precision more than 12 percentage points higher, but for the model with $k = 100$, it was only 0.29 percentage points.
Nonetheless, we still recommend using it to ensure more generalizable results.
Furthermore, there are sometimes more efficient techniques to implement cross-validation for specific models (see, e.g., Cheng et al.~\cite{cheng2017efficient} and Cawley et al.~\cite{cawley2008efficient}).
In our case, the more efficient implementation of the \textit{leave-one-out validation} for the k-nearest-neighbor model was more than 12 times faster.
Furthermore, cross-validation via random splits always achieves different results.
In some of our runs, precision could differ by up to 5 percentage points just by using different training and test data sets.
As a consequence, it was critical to train and test the model several times and average the results.

\textbf{Continuously measure model quality, performance, and drift:}
During fine-tuning of the ML models, we continuously measured the model prediction quality via precision to identify how the many new versions compared.
This was especially important after drastically changing the feature selection.
Many of the models we used had a strong bias, e.g., due to unbalanced training data.
Once we realized the potential effect of new data on prediction quality, we ensured that newly trained models were only incorporated into the system if model quality was better than for the previous version.

The final version of the system makes approximately 760 predictions each day before the stock market opens, which took, on average, 39 seconds.
This duration was important to know for deciding when the system should start its decision-making process.
For example, we could configure the system to start 10 min before the stock market opens.
However, if the ML component is changed or more data is used, this process could take longer, meaning that the system would lack decisions for investing.
To avoid this, we continuously measured the prediction duration of the current model.

We did not discover any serious model drift during development.
However, we assumed that the state of the stock market would impact model precision.
To verify this, we tested one model purely on data from September 2008, since this was the month the stock market crashed.
The chosen model predicts if a stock will increase by more than 2.6\% during the next day.
This model had a precision of 63\% when tested on random data, but only scored 59\% for September 2008.
This indicates that the state of the market can have an impact on prediction quality.
While such crashes are rare, there are other events that also have an impact, e.g., the war in Ukraine, which caused stocks like Rheinmetall to increase and others like the EPAM Systems stock to decrease massively~\cite{ukraine}.

\textit{Analysis:}
In our case, continuously measuring model quality was essential to select the best versions, measuring prediction duration was critical to ensure system correctness, and measuring model drift was important to ensure that system effectiveness does not unexpectedly decrease.
In summary, applying this practice increased observability and improved the development process and resulting system.

\textbf{Review model training scripts:}
At first, our training script was so short and simple that reviewing it was not necessary.
However, later on, we developed three different training scripts for two ML algorithms, which led to considerably more complexity.
During a review of this code, we identified and implemented an optimization procedure which increased model precision from 59\% to 61\%.
However, this improvement also introduced two bugs.
One of them, an infinite loop bug, was discovered because training took much longer than expected.
As a result, we performed another review of the training code, and found and fixed the bug within 5 min.
Although this bug was critical for continuing development, we only conducted the review due to our suspicion, and should have done so much earlier.
The second bug, a floating-point error, was only detected because the newly added optimization process did not show any effect.
Python returns $0.010999...$ as the result for $1.011-1$, which we discovered by debugging the newly added code.
Afterwards, we reviewed the entire training script again, and discovered several ways to improve understandability, such as renaming variables or extracting methods.

The second training script we implemented was a k-nearest-neighbor algorithm, and a first review of it did not reveal any bugs.
However, it again revealed several ways to increase code understandability.
Furthermore, we conducted a second review with the external ML engineer, who pointed out several ways in which the training script could be further improved, e.g., by using NumPy\footnote{\url{https://numpy.org}} to achieve greater performance and modifying the optimization procedure to achieve a better tuned model.
We used his proposal to develop another k-nearest-neighbor training script, which, however, also contained a bug related to element removal based on the list index.
This bug was not identified during the conducted review.
Only after testing the training script, we finally discovered and removed it.

\textit{Analysis:}
Conducting code reviews is already a well-established practice, as it improves software quality~\cite{mcintosh2016empirical}.
Our case exemplifies that this is also valuable for ML training code.
By conducting such reviews, we were able to detect and fix bugs and to improve code quality.
Especially for complex training scripts, some bugs can be difficult to detect.
The review with the ML engineer led to further improvements of the training script.
We conclude that conducting such reviews was an important practice for us.
However, one should be aware that one review might not be enough.
Fixing detected bugs can unintentionally lead to new bugs.
Therefore, we strongly recommend that, after fixing bugs or modifying the training script, another review should be conducted.

\textbf{Test all feature extraction code:}
During ML component development, most time was spent on feature engineering code.
As this code also changed quickly during exploratory model development, it was difficult to decide when we should write tests for it.
This code was also fairly complex, e.g., our first feature extraction script parsed thousands of CSV lines just to create a single data point.
Creating dummy test data would therefore have been very time-consuming.
Instead, we conducted manual tests by comparing the extracted features with the corresponding data.
The first bug we discovered was caused by two duplicate lines that added new data points, which led to profitable data points having different feature vectors than unprofitable ones and an accuracy of 99.5\%.
Out of suspicion, we decided to test the feature extraction code and added logging, which quickly led to the discovery and fixing of the bug.
Once the feature extraction code became more stable, we also added equivalence class testing to ensure a reasonably broad coverage of the training set.
For each equivalence class, we compared one of the extracted features with the corresponding stock data, which led to the discovery of another bug for one class.

\textit{Analysis:}
We draw several lessons from this experience.
First, testing the feature extraction code is critical to ensure the prediction quality of the system.
Second, if we had tested this code more frequently, we would have saved several days of work.
Third, it should raise suspicions when the model suddenly performs unexpectedly well.
Verify that this is not caused by any bugs.
Fourth, a bug might only subtly corrupt a fraction of the data.
Therefore, ensure that the conducted tests cover the majority of the data set, e.g., by using equivalence class testing.
Overall, this practice was very effective, and we should have used it earlier and more extensively.

\textbf{Automate hyperparameter optimization and model selection:}
Automating the hyperparameter optimization always requires additional code.
When we only had a small set of values for the hyperparameters, writing this code could take more time than just manually trying out each combination.
However, once the model training took several minutes, it was always more effective to automate this optimization, since it allowed us to simultaneously work on something else.
Furthermore, there was a trade-off between training duration and model prediction quality, as extensive automated optimization increased both.
At some point, we therefore executed the hyperparameter optimization during the night.
Although this required more resources and planning, the precision of the tuned models was always at least 0.5\% better (usually more) than with a lightweight optimization taking less than one hour.
In most cases, we used a particle swarm optimization algorithm implemented in the tool Optunity\footnote{\url{https://optunity.readthedocs.io}}.
In some cases, we also used a conventional grid search algorithm to compare the results.

Additionally, we experienced that the hyperparameter optimization duration could be drastically decreased with more efficient model implementations.
Our final ML component used 19 individual k-nearest-neighbor models, for which the hyperparameter optimization should find the optimal $k$ in the range from 5 to 50, i.e., a simple grid search.
Switching to a model implementation based on NumPy reduced the optimization process of the combined 19 models from 52 min to only 3 min.
Alternatively, this increased performance could be used to execute a more extensive optimization process in less or comparable time.

\textit{Analysis:}
Automating the hyperparameter optimization definitely had a positive effect on our process, since it improved model precision by more than 4\% in some cases.
Even though manual fine-tuning could be faster for less complex configurations, automating hyperparameter optimization was an essential step to increase our model precision.
Additionally, it also allowed parallel development on other tasks.
Still, extensive optimization requires a lot of time and resources, in our case more than 5 hours, which needs to be planned accordingly.
More efficient model implementations can be a way to address this.

\textbf{Log prediction results together with model version and input data:}
Since we used between 20 and 65 features in the beginning, logging predictions was not valuable, as it was difficult to understand why model predictions were incorrect.
However, at some point, we wanted to identify which additional features increase model precision.
We therefore enabled logging and analyzed the feature values and precision with charts, which led to a feature set that increased model precision from 62\% to 66\%.
However, the real value of logging during training became only apparent when we used a set of models instead of only one, which also required adding the model version.
The deployed system makes approximately 50,000 predictions every day.
Logging all of them would create around 12 GB each year.
A more efficient approach could only log positive predictions that the system actually invests in.
This would reduce the log data size to less than 120 MB per year.

\textit{Analysis:}
During the development, logging predictions was an essential step to generate insights into how the model can be further improved.
To improve the model after deployment, logging predictions in production can definitely be valuable, too.
However, one should consider the required space, and provision a feasible infrastructure for this.
More efficient partial logging can be an alternative.

\textbf{Collaborate with multidisciplinary stakeholders:}
We conducted three meetings with two external stakeholders.
The first meeting was with a professional stockbroker and took 44 min, during which we reviewed the completed system design.
Furthermore, we clarified domain-specific questions that came up during the requirements analysis and system design. 
During the discussion, additional questions emerged, to which the stockbroker provided valuable domain-specific knowledge.
For example, one unexpected result was the necessary precision for the ML models.
We had expected something around 90\%, but the stockbroker explained that, even for the best traders, only 55-60\% of investments are profitable.
This changed our way of thinking: instead of building a system that rarely makes incorrect predictions, we focused on a precision of at least 60\%, but additionally tried to reduce money lost on unprofitable investments and to increase money gained for the inverse case.
This changed the system requirements and impacted the system design.

The second meeting was held with an ML engineer and took 54 min, with the goal to improve the system design.
We received a substantial amount of valuable feedback for different stages of the ML development process, such as model selection, feature selection, hyperparameter optimization, and deployment.
For example, he suggested calculating the feature importance for each feature to identify useless features and using Optunity\footnote{\url{https://optunity.readthedocs.io}} to automate the hyperparameter optimization.
The last meeting with the same ML engineer took 70 min, during which the finished system prototype was reviewed.
Again, several valuable improvements were suggested, such as using NumPy to improve model performance.

\textit{Analysis:}
Applying this practice highlighted the importance of acquiring missing domain knowledge for ML projects.
If developers lack this knowledge, it is essential to collaborate with domain experts.
For example, although we were able to develop a model with a precision of 83\%, the remaining 17\% of unprofitable investments would still result in a net loss for the system.
Based on the stockbroker feedback, we aborted this misguided strategy and were able to increase system gains.
According to \citet{piorkowski2021ai}, knowledge in a domain as complex as AI cannot be transferred during a single meeting, as new questions will continuously arise during development.
We experienced the same after the meeting with the stock market expert, which highlights the importance of continuous collaboration with domain experts.

The technical meetings with the ML engineer improved system performance and effectiveness, i.e., incorporating reviews with an expert more experienced with certain ML aspects was extremely helpful.
For example, the engineer pointed out improvements which resulted in more than 20 times faster model performance and 1.5\% more model precision.
Overall, this practice was very effective for our case study.

\subsection{Experienced Challenges (RQ2) and Solutions (RQ3)}
In this section, we describe challenges experienced during the development and also how we addressed them (\textit{solution}).
Table~\ref{table:challenges} provides an overview of these challenges.

\textbf{Strong influence of ML implementation on system design:}
One fundamental design question at the beginning revolved around how the system selects stocks to invest in.
Initially, we thought about prefiltering stocks, e.g., based on strong price increases, to reduce the number of necessary predictions.
Moreover, we were unsure about the concrete stock data to be used for the predictions.
Intraday data seemed the most promising, but during the development stage, none of the models trained with this data had sufficient precision, which made us consider daily stock data as well.
Many of these questions strongly influenced the architecture and workflow of the system.
For example, if the model only uses daily stock data, predictions can be made before the stock market opens.
However, with intraday data, it would be essential that predictions finish in a matter of seconds.
At first, we designed and implemented the overall system based on our initial idea of the best way, and only then started to develop the ML component.
This turned out to be a huge mistake because the overall system often had to be changed profoundly whenever we found a better way to predict stocks.

\textit{Solution:}
Once we realized this strong influence, we focused on the ML component to detect the most effective way of stock prediction by implementing and comparing a variety of ML models with different investment strategies.
Only after we had discovered the most effective strategy (use daily stock data; only consider stocks that increased by at least 10\% the day before; buy stocks as soon as the market opens; sell stocks as soon as the predicted profit, a predetermined loss, or the end of the day is reached), we then changed the overall system design and workflow to match this strategy.

\textbf{Finding effective data sources for the deployed system:}
At runtime, the system needs to query up-to-date stock prices to make predictions.
However, most APIs offering such data have request limits, which would be especially hindering for using intraday data.
To avoid this, we initially used the \texttt{lemon.markets} API\footnote{\url{https://www.lemon.markets}}, since it allowed 200 requests per minute with a free account.
During development, however, \texttt{lemon.markets} changed this to only 10 requests per minute, which severely limited our system's effectiveness.
This issue is related to \textit{unstable data dependencies}, which \citet{Sculley2015} described as a facet of data dependency debt.

\textit{Solution:}
Initially, we tried restricting the system to only a small set of stocks, which were selected every day based on past increases.
However, we later discovered the Alpaca API\footnote{\url{https://alpaca.markets}}, which offers similar functionality but allows 1,000 requests per minute.
Because of this enormous advantage, we switched to this new API, even though changing the respective component required some efforts.
Using several parallel requests for 1,000 stocks each, we reduced the average time to retrieve all available stock data to 8.7 seconds.

\textbf{Finding effective data sources for ML training:}
While the Alpaca API was great for retrieving recent stock prices at runtime, the requirements for the historic ML training data were very different.
In a first version, we assumed that using historic intraday stock data in one-minute intervals would be most effective.
Because such data has an enormous size, it was difficult to find an effective data source.
Initial searches revealed a few APIs that offered such historical intraday stock data, but often with unsuitable quality or strict request limiting.

\textit{Solution:}
To find a suitable source, we therefore conducted a more systematic search, which revealed that the most popular API for this purpose is the Alpha Vantage API\footnote{\url{https://www.alphavantage.co}}~\cite{alphavantagerecommendation1,alphavantagerecommendation2}.
In comparison to other options, it offered the best data quality, history, and granularity.

\textbf{Efficiently retrieving the ML training data:}
The Alpha Vantage API offers stock data for the last 720 days.
Data for one stock is split into 24 CSV files, each representing 30 days, and the API allows 500 requests, i.e., files, per day.
Retrieving the data for all 4,927 Nasdaq stocks would therefore take 237 days.
The API also has a limit of 5 requests per minute.

\textit{Solution:}
Due to these restrictions, we decided to focus only on stocks with a market capitalization of at least two billion dollars.
In the Nasdaq, these are 707 stocks.
Retrieving their data required 16,968 API requests, i.e., it took 34 days of 500 requests.
Additionally, the retrieval script needed to adhere to the five requests per minute limit by waiting long enough between requests.
After receiving the data for around 400 stocks, we decided to first focus on developing an effective model before retrieving more.

\textbf{Time-consuming feature extraction:}
When we still trained our models using intraday data, the training data size was 3.57 GB.
Extracting features took between 3 and 18 min, which massively slowed down the tuning process.
Since we frequently changed features during the tuning, storing extracted features would not solve this issue.

\textit{Solution:}
As a workaround, we decided to only use the data for 10 stocks, for which feature extraction took less than a minute.
Once the tuning process with this small data set led to a satisfactory model, we then started testing the model on a larger set of data.
Afterwards, we statically stored the extracted features, as loading them takes less than one second.
This solved the problem completely, but is only effective when feature selection is stable.
When we switched to daily data later on, feature extraction time reduced to an average of 14 seconds, which also was no longer an issue.

\textbf{Data quality issues:}
Many retrieved intraday files contained gaps.
In most cases, only a couple of minutes were missing, but gaps could sometimes reach multiple hours, days, or in rare extreme cases months.
It was difficult to estimate the negative impact of these gaps, which made it challenging to decide on a reasonable threshold.
Writing a sophisticated script to check all files also requires additional effort.

\textit{Solution:}
An autonomous system requires a base-level of automatic data quality assurance.
We therefore implemented an automated quality check procedure.
To avoid an overly complex solution and long runtime, we used the number of lines per file as a metric.
If a file contained fewer than 100 lines, the complete associated stock data would be removed from the training set.
Choosing a simple solution turned out to be the right decision, since we later switched to daily data anyway.
More efforts for improving data quality before starting model development would have wasted time.
Early on, we recommend to first develop a satisfactory model on a small set of high-quality data before spending substantial time to improve overall data quality.

\textbf{Managing a large number of files:}
We received a total of 9,888 files for 412 stocks, which was hard to manage and also made the training process unnecessarily complex.
The training script would be much simpler if there was one file per stock.

\textit{Solution:}
We therefore modified the script for stock data retrieval to perform this merge for each received stock.
This required a bit of additional work, especially since reordering file contents was necessary.
Nonetheless, this massively simplified the feature extraction code, thereby improving code understandability and simplifying the remaining development.

\textbf{Selecting an effective ML algorithm:}
An initial survey of the literature revealed that many authors propose to use support vector machines (SVM) to predict stock movements~\cite{svm1,svm2,svm3}.
However, this type of model performed poorly with our data.
We therefore started manually testing many different models with different data sets (intraday or daily data), but were initially unable to identify a clear winner.

\textit{Solution:}
At some point, we decided to focus on models using daily data, since many authors proposed this~\cite{dailyrecom1,dailyrecom2,dailyrecom3}.
This reduced the suitable models to choose from.
Furthermore, we created an automated model selection method to detect the model with the highest precision.
Initially, decision trees achieved the best results, but with changing feature selection, other models performed best.
At the end, the model selection method showed that k-nearest-neighbor had the best precision, so we focused on improving this type of model even more.

\textbf{Selecting effective model features:}
The most challenging task during ML component development was deciding what features best indicate upward stock movement.
There are many possibilities to choose from, and we lacked the stock trading expertise a professional trader has.

\textit{Solution:}
To solve this, we tested models using a wide variety of features, often many in parallel.
We then calculated and visualized the importance of each feature, which informed further changes to the feature selection.
However, since none of the models performed as well as expected, we additionally surveyed the literature for similar ML models, and tried out these solutions.
The best precision was achieved using the features suggested by \citet{vijh2020stock}:

\begin{enumerate}
    \item Stock High minus Low price $(H-L)$
    \item Stock Close minus Open price $(C-O)$
    \item Stock price 7 days moving average
    \item Stock price 14 days moving average
    \item Stock price 21 days moving average
    \item Stock price standard deviation for the past 7 days
\end{enumerate}

Furthermore, we added the stock volume as another feature, since this increased model precision even further.
During testing different features variations, we also discovered that the model can be improved by altering the first two features to $(H/L) - 1$ and $(C/O) - 1$ respectively, thereby modeling the relative difference rather than the absolute difference.
This increased precision by another 2\%.
The combination of surveying the literature and calculating and visualizing feature importance was very effective for us.

\section{Discussion}
From the 10 applied \textbf{practices}, the majority had a positive influence on our system and development process.
One exception was \textit{Capture the training objective in a metric that is easy to measure and understand}, as we were not really able to capitalize on this practice.
While trying to apply it led to thinking about our system, its objective, and investment strategies, we were ultimately not able to define a goal that was both measurable and understandable.
Another practice with only a small effect was \textit{Log prediction results together with model version and input data}.
Applying it often did not generate a lot of helpful insights.
Only when aiming to improve feature selection, it briefly became valuable.

\begin{table}[ht]
\caption{Applied practices and their perceived effects ($--$, $-$, $0$, $+$, $++$)}
\label{table:practice-summary}
\centering
\begin{tabular}{p{7.5cm}l}
Practice & Impact \\
\hline
\hline
Collaborate with multidisciplinary stakeholders & $++$  \\
Standardize and automate data quality check procedures & $+$  \\
Use error validation and categorization & $+$  \\
Use cross-validation & $+$  \\
Continuously measure model quality, performance, and drift & $+$  \\
Review model training scripts & $+$  \\
Test all feature extraction code & $+$  \\
Automate hyperparameter optimization and model selection & $+$  \\
Log prediction results together with model version and input data & $0$  \\
Capture the training objective in a metric that is easy to measure and understand & $-$ \\
\end{tabular}
\end{table}

\begin{table}[ht]
\caption{Challenges experienced during the case study (overall perceived difficulty can be low, medium, or high)}
\label{table:challenges}
\centering
\begin{tabular}{ll}
Challenge & Difficulty \\
\hline
\hline
Selecting an effective ML algorithm                    & high   \\
Selecting effective model features                     & high   \\
Finding effective data sources for ML training         & high   \\
Efficiently retrieving the ML training data            & medium \\
Finding effective data sources for the deployed system & medium \\
Data quality issues                                    & medium \\
Strong influence of ML implementation on system design & low    \\
Time-consuming feature extraction                      & low    \\
Managing a large number of files                       & low    \\
\end{tabular}
\end{table}

While most other practices definitely improved the system (see Table~\ref{table:practice-summary}), some of them were also fairly situational, e.g., \textit{Continuously measure model quality, performance, and drift}.
The real value of this practice would probably only become apparent when operating the system for several months, which we did not do.
Additionally, applying some practices would most likely have been different with a larger team of developers, e.g., \textit{Review model training scripts}.
Despite most practices being helpful, the most effective one was by far \textit{Collaborate with multidisciplinary stakeholders}.
Incorporating both a domain and ML expert into the process led to a multitude of improvements that would most likely have been impossible without this practice.

Overall, the state of AI engineering practices seems to slowly mature towards effective solutions.
However, one small issue is that applying several practices may not be straightforward enough for most practitioners.
Some practices will be easier to apply with a strong software engineering background, others with an ML or data science one.
But without more concrete steps or specific tool support, it may take time to successfully apply a practice, especially for a novice team.
We hope that our experiences can partly mitigate this issue.

Regarding our encountered \textbf{challenges}, most of them – and also the ones perceived as most critical – were related to acquiring suitable data and training an effective ML model for the use case at hand (see Table~\ref{table:challenges}).
Since we primarily have an SE background and low domain familiarity, this is not really surprising.
Nonetheless, nearly all of these challenges have occurred in literature before.
\citet{ibm} and \citet{Sculley2015} mentioned the challenge of unreliable data sources, while \citet{zhou2017machine} described efficiently receiving data as challenging and potentially costly.
Regarding the strong influence of the ML implementation on system architecture and workflow, \citet{Sculley2015} described the related CACE principle: \enquote{Changing Anything Changes Everything}.
While focusing on the ML model for some time was an effective workaround, we probably should have used more encapsulation via custom interfaces to isolate changes.
\citet{l2017machine} also named a number of our challenges, e.g., high data volume, data processing performance, and data quality issues, with solutions similar to ours.
Lastly, feature~\cite{zhou2017machine,l2017machine} and model selection~\cite{monteiro2021meta} were also described as challenging.
However, most authors propose to simply use deep learning as a solution, which may lead to several new challenges.
In our case, surveying the literature, analyzing feature importance, and automating model selection were sufficiently effective.

\section{Threats to Validity}
Some limitations have to be mentioned for our results.
First, we need to be careful when generalizing from a single, rather \enquote{synthetic} case study to other contexts.
As already mentioned, a different development process and larger team size (instead of one developer), but also a different domain may likely lead to different results.
For example, it might be less effective to extensively collaborate with a domain expert in less complicated domains.
Nonetheless, we assume that most practices used in this study will also be helpful elsewhere, albeit with potentially different application and impact.
Furthermore, we assume that other challenges may occur, e.g., communication challenges with a larger team.
However, many of our experienced challenges should be fairly universal, regardless of the process or domain.

Since evidence collection and analysis was qualitative, there is the potential for subjective bias.
The effect of practices or the difficulty of challenges could definitely be perceived differently by people with other backgrounds or experience.
Nonetheless, we believe our results add valid, contextual evidence for AI engineering practices and challenges.
Still, more research is needed to collect data in different domains and from other developers.

\section{Conclusion}
To collect evidence on the effectiveness and applicability of AI engineering practices, we conducted a case study, during which we developed an autonomous stock trading system.
Via field notes, we documented how 10 selected practices were applied, as well as challenges we encountered and their solutions.
Overall, the majority of practices improved the system and development process, especially the practice \textit{Collaborate with multidisciplinary stakeholders}.
However, a few practices showed less effects, e.g., \textit{Capture the training objective in a metric that is easy to measure and understand}.
We also encountered several challenges that were mentioned in literature, and were able to overcome them, sometimes in ways not described before.
Overall, our results provide rich, qualitative evidence that popular AI engineering practices are mostly effective, but that applying them is sometimes not straightforward.
While our results can provide some guidance and experience in this regard, future research should analyze how these practices could be made more actionable for practitioners, especially novices.
To support such endeavors and other AI engineering research, we publicly share the system\footnote{\url{https://github.com/Marcel0503/Autonomous-Stock-Trading-System}} and the study artifacts\footnote{\url{https://doi.org/10.5281/zenodo.7566146}}.

\section*{Acknowledgment}
We kindly thank Markus Böbel (NorCom AG) and the financial analyst (who preferred to remain anonymous) for acting as external stakeholders to apply the practice \textit{Collaborate with multidisciplinary stakeholders}. This research was partially funded by the Ministry of Science, Research, and the Arts (MWK) Baden-Württemberg, Germany, within the Artificial Intelligence Software Academy (AISA).

\bibliographystyle{IEEEtranN}
\bibliography{references}

\end{document}